\documentclass[twocolumn,aps,fleqn]{revtex4-1}
\usepackage{graphicx}
\usepackage{amssymb}
\usepackage{amsfonts}
\usepackage{amsmath}
\usepackage{amsbsy}
\usepackage{natbib}
\usepackage{comment}
\usepackage{color}
\usepackage{float}
\usepackage{array}

\newcommand{\be}{\begin{equation}}
\newcommand{\ee}{\end{equation}}
\newcommand{\ben}{\begin{eqnarray}}
\newcommand{\een}{\end{eqnarray}}

\begin{document}

\title{Electrically driven  magnetic antenna based on multiferroic composites}

\author{X.-G. Wang$^{1}$}

\author{A. Sukhov$^{1}$}

\author{L. Chotorlishvili$^{1}$}

\author{C.-L. Jia$^{3}$}

\author{G.-H. Guo$^2$}

\author{J. Berakdar$^{1}$}
\email{Jamal.Berakdar@physik.uni-halle.de}

\affiliation{
$^{1}$Institut f\"ur Physik, Martin-Luther Universit\"at Halle-Wittenberg, 06099 Halle (Saale), Germany\\
$^{2}$School of Physics and Electronics, Central South University, Changsha 410083, China \\
$^{3}$Key Laboratory for Magnetism and Magnetic Materials of the Ministry of Education, Lanzhou University, Lanzhou 730000, China
}

\begin{abstract}
We suggest and demonstrate via large scale numerical simulations an electrically operated  spin-wave inducer based on  composite multiferroic junctions. Specifically, we  consider an interfacially coupled ferromagnetic/ferroelectric structure
that emits controllably spin waves in the ferromagnets if the ferroelectric polarization is poled by an external electric field. The roles of geometry and material properties are discussed.
\end{abstract}

\date{\today}

\maketitle
%%%%%%%%%%%%%%%%%%%%%%%%%%%%%%%%%%%%%%%%%%%%%%%%%%%%%%%%%%%%%

\section{Introduction}

Spin wave excitations and manipulation is an important subfield of spintronics dubbed magnonic, which is also of relevance to spin wave-based computing and communication devices \cite{ChUm15, LeNk11, KrUg10, SeRg10}. To effectively excite and manipulate the magnonic spin current, various methods are proposed and experimentally tested, such as microwave field and spin polarized current \cite{SeRg10, KiSe03, KyUn04}. However, high power dissipation and scaling issues are serious drawbacks. Recent electric-field induced spin wave generation using multiferroic systems points to promising directions \cite{ChEr14}.

Multiferroics \cite{FiEb05, EeMa06, RaSp07, VeJa11} may have multiple coupled ferroic orders  such as  ferroelectric (FE), and/or ferroelastic, and/or ferromagnetic (FM) orders and hence may respond to a variety of external probes which opens the way for qualitatively new  applications.
Here we will be dealing with magnetoelectric composites, i.e. FE/FM heterostructures \cite{NoSh12, VaCr11, PaGo12, Vaz12}. In principle the underlying idea includes also strain-based magnetoelectrically coupled structures, albeit detailed numerical simulations as those presented here are necessary to estimate the relevance of the associated effects.
Generally, composite FE/FM are most interesting, as in these materials the interfacial magnetoelectric coupling turned out to be quite sizable even at room temperature, as evidenced by
  numerous  studies   \cite{DuJa06, RoSt08, CaJu09,JiWe14}. Experimental verification and a direct observation of the ME coupling  was reported  recently  for Co/BaTiO$_3$-interfaces \cite{JeBa13} and Co$_{92}$Zr${_8}$/PMN-PT-films \cite{JiWa15}.
 Several coupling mechanisms are discussed in the literatures. For example,  for strain-induced magnetoelectric (ME) coupled FE/FM structure, the strain due to e.g. lattices mismatch between the FE and FM can strongly influence the magnetic properties like magnetic anisotropy in the FM layer \cite{VaCr11, SaPo07, JiSu12}.
  For charge-mediated ME coupling when the FM is metallic, the uncompensated charge at the edge of the FE are  balanced by the itinerant electrons in the
 FM, which in turn are spin-polarized and also coupled via the exchange interaction to the net FM magnetization \cite{DuJa06, MeKl11, JiWe14, JiWa15}. Generally, several coupling mechanisms are operational at the same time. Via experimental arrangement one may gain insight into their relative strengths. For instance, in the experiments on polycrystalline, several-dozens-nm thick Co deposited on single crystal BaTiO$_3$ \cite{JeBa13}, the main contribution is shown to be  provided by the charge-mediated ME coupling (similar findings have been reported for multiferroic Co/P VinyliDene Fluoride-TriFluoroEthylene \cite{MaDu11}).

In this paper we will simply assume the existence of the ME coupling (using previous knowledge) and study the consequences therefore, such as spin wave emission via electric stimulations. In turn the properties of these spin waves carry some footprints of the underlying ME coupling.
  We assume that an external electric field induced a FE polarization dynamics
   in a FE attached to a FM nano stripe which leads thus to  spin wave excitation. By varying the shape of the multiferroic antenna, one can manipulate the spatial distributions and intensities of excited spin waves.

\section{Origin of the quenched magnetization}

To be specific, we employ the results of Ref. \cite{MeKl11}, where layers of BaTiO$_3$ were grown on 3MLs of Fe(001). The authors report on two remarkable observations: the magnetic moment is induced in the initially non-magnetic TiO$_2$ layer closest to Fe and when the FE polarization is switched, the induced magnetic moment changes by approximately $\Delta \mu_{\mathrm{TiO}_2}=0.4\mu_{\mathrm{B}}$.
We start with Figure 4 of Ref.\cite{MeKl11}, which shows the ab-initio-calculations of the layer-resolved magnetic moments (red symbols).
Assuming the exchange interaction to be a direct one and ferromagnetic between the induced and the magnetic moments in the first monolayer of iron from the interface, one can estimate the total change of the interfacial magnetization as $\Delta M_{\mathrm{TiO}_2}(P_{\uparrow}-P_{\downarrow})=\Delta \mu_{\mathrm{TiO}_2}/V_{\mathrm{TiO}_2} \approx 0.46\cdot 10^6$ A/m, where $V_{\mathrm{TiO}_2}=(2\cdot 10^{-10})^3$ m$^3$. Now from Figure 3 of Ref.\cite{MeKl11} we estimate the quenched length near the Fe-interface to be $d\approx2A$. This allows estimating  (the relative to the saturation magnetization at $T=0$ K) the quench of the interfacial magnetization to be around $\Delta M_{\mathrm{TiO}_2}(P_{\uparrow}-P_{\downarrow})/M_{\mathrm{S Fe}}=27\%$. Switching of the FE polarization quenches the interfacial magnetization and excites so propagating spin waves.

Though the experiment \cite{MeKl11} provides important quantitative estimates regarding the strength of the magnetoelectric effect, it contains little or no information on the spatial alignment of the magnetization near the interface. To the best of our knowledge, the authors are not aware of any experimental evidence for the FM order in the vicinity of the interface for such a structure. Nevertheless, in the recent study \cite{JiWe14} a theory elucidating the spin alignment near the FE/FM-interface was proposed. According to the suggested mechanism, an interfacial spiral spin density extending over the spin diffusion length in the ferromagnet functionalizes the interface for the electric control. Taking Fe/BaTiO$_3$-interface as an example with the spin diffusion length around 8 nm in Fe \cite{BaPr07}, we conclude that the spin spiral ordering is highly localized when thinking of FM extensions up to several hundreds nanometers.

%\begin{widetext}
\begin{figure}[h]
\centering
\includegraphics[width=0.49\textwidth]{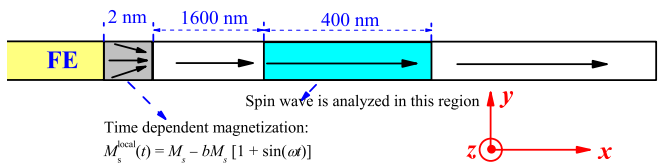}
\caption{Schematics of the considered two-dimensional $3000\times 30 \times 10$nm$^{3}$-structure. The area corresponding to the magnetization quench due to the magnetoelectric coupling is shown on the left ($2$ nm-thick, grey color). The area where the induced spin waves are analyzed is in the middle (400nm-area).}
\label{fig_1}
\end{figure}
%\end{widetext}

\section{Results of numerical simulations}

We first consider a magnetic nanostripe with $3000$ nm in length (along the x-direction), $30$nm width and with $10$nm thickness (Fig. \ref{fig_1}). The spin wave is excited at the left boundary, since the contact of the FM with the FE driven by an E-field leads to a quenched magnetization. In principle we can also model
 the FE dynamics in the presence of the
  strong driving E-field, as done in our previous studies, for the purpose of present study this doing is not necessary (the spin dynamics is much slower than the interfacial transient electric dipole switching). Poling the FE part  results  in a quenched magnetization at the boundary in the form $M_{\mathrm{S}} ^{\mathrm{local}}(t)=M_{\mathrm{S}} - b M_{\mathrm{S}} [1 + \sin(\omega t)$], where $\omega/(2\pi)=50$ GHz is taken ($b$ is a constant).

We note that the parameter $b$ characterizes the strength of the quench of the magnetization that occurs at the ferroelectric-ferromagnetic interface due to the switching of the ferroelectric polarization. Thus, the upper limit of $b$ is $b=\Delta M_{TiO_{2}}\big(P_{\uparrow}-P_{\downarrow}\big)/2M_{sFe}$. The angular frequency of the quenched magnetization is defined by the frequency of the time dependent electric field that switches the ferroelectric polarization $P_{\uparrow}\leftrightarrow P_{\downarrow}$.

In general the direction of the quenched magnetization is determined by demagnetizing fields which give rise to slightly tilted magnetic moments at the boundary. In the following, to examine the influence of dynamics in the quenched region, we contrast two situations: when the amplitude of the saturation magnetization is steered, a) fixed quench case: direction of the magnetization vector is fixed and b) free quenched case: direction of the magnetization vector is free. Experimentally interesting is to consider the relevant  and realistic case of  a free quenched magnetization. A fixed quench case delivers information on whether a fixed quench of magnetization can excite spin waves or not.
Further on, we will analyze the spin-wave spectrum in the $400$nm-area (Fig. \ref{fig_1}), which is enough far away from both edges to exclude boundaries effects.

The modeling proceeds within the micromagnetic framework using the \texttt{mumax3}-simulation package \cite{VaLe14} with the cell size $2\times 2 \times 10$ nm$^3$ and the material parameters related to the bulk iron: the saturation magnetization $M_{\mathrm{S}}=1.7\cdot 10^6$ A/m, the exchange stiffness constant $A=2.3\cdot 10^{-11}$ J/m, the anisotropy constant $K=4.8\cdot 10^4$ J/m$^3$ with the easy axis aligned along the \textit{x}-direction \cite{Coey10} (since spin waves are low energy magnonic excitations, we neglect the second constant $K_2$ of the (cubic) anisotropy). The damping parameter is taken as a maximum value calculated via \textit{ab-initio} $\alpha=0.01$ \cite{GiId07}.

To assess and remove possible numerical spurious effects, such as the choice of the initial state and boundary conditions, we start with the randomly chosen magnetic state. However, we propagate  the magnetization within the time interval $\tau _{\mathrm{sim}}=200$~ns that exceeds the relaxation time of the homogeneously magnetized FM $\tau_{\mathrm{rel}}(\mathrm{Fe})=\tau^{\mathrm{prec}}/\alpha=\frac{\pi}{\alpha}\frac{M_{\mathrm{S}}}{\gamma K} \approx 63$~ns.
Then the Fourier analysis is performed only for the last $40$~ns and this naturally excludes the memory effect related to the initial state.

The frequency domain relevant to the spin wave can be roughly estimated from the energy gap in the dispersion relation. A precise expression for the finite size bulk system reads \cite{ChOi07,GuSl03}:
\setlength{\mathindent}{0cm}
\begin{equation}
\begin{split}
&&\displaystyle w_{sw}^{2} = \left[\omega_{H} + a \omega_{m} k_{n}^{2} + \omega_{m} P(k_{n} t) k_{y,n}^{2} / k_{n}^{2}\right] \\
	 &&\times \left \{ \omega_{H} + a \omega_{m} k_{n}^{2} + \omega_{m} \left[1 - P(k_{n} t) \right]\right \}.
\end{split}
\label{eq_1}
\end{equation}
Here $P(k_{n} t) = 1 - \left[1-\mathrm{exp}(-k_{n} t)\right]/\left(k_{n} t\right)$  and we introduce the following notations: $\omega_{H} = \gamma H_{in}$, where $H_{in}$ is the internal magnetic field consisting from the magnetocrystalline anisotropy and static demagnetization fields, $\omega_{m}=\gamma M_{s}$, $a=2 A/(\mu_0 M_{s}^{2})$ and the spin-wave vector $k_{n}^{2}=k_{x}^{2}+k_{y,n}^{2}$ is quantized along the \textit{y} axis $k_{y,n}= (n + 1) \pi / w_{d}$. For the dipolar pinning boundary condition \cite{KYGu05}, $w_{d}$ is supposed to be $w_{d}=w d(p) /[d(p)-2]$, where $d(p) = 2 \pi / \{p [1 + 2\mathrm{ln}(1/p) ]\}$ is the effective pinning parameter, $p = t / w$ and $t$, $w$ are thickness and width of the nano stripe. The cutoff frequency is of the order   $\omega_{cut}=32$ GHz. In the following, we will consider frequencies above this threshold value.

\begin{figure}[h]
\vspace{2ex}
\centering
\includegraphics[width=0.49\textwidth]{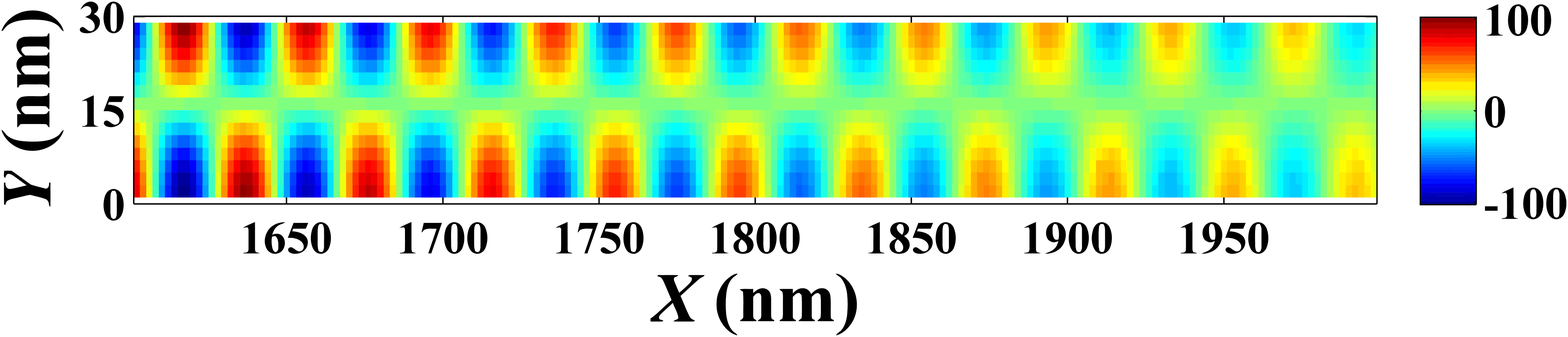}
\vspace{0ex}
\includegraphics[width=0.49\textwidth]{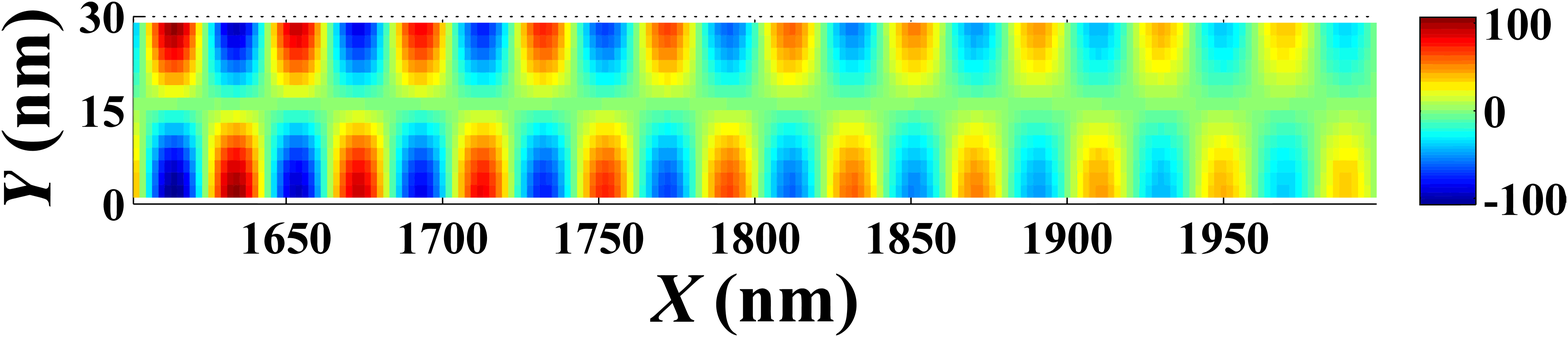}
\caption{Spatial distribution of the excited spin wave ($n=1$ mode) of the $m_{\mathrm{z}}$-magnetization component for the fixed (panel above) and free (panel below) quenched magnetization direction ($\omega/(2\pi)=50$ GHz, temperature $T=0$ K, external field $B=0$ T). The modulus of the quenched magnetization alters according $M_{\mathrm{S}} ^{\mathrm{local}}(t)=0.94M_{\mathrm{S}}-0.06M_{\mathrm{S}}\sin(\omega t)$. Unit of the spin-wave profile is A/m.}
\label{fig_2}
\end{figure}

In our simulations time-dependent magnetization is steered homogeneously along the \textit{y}-direction in the vicinity of the FE/FM interface. This mimics cooperative effect of the ME coupling and FE polarization driven via the external electric field. Equilibrium spin wave profile  $400$nm in length (Fig. \ref{fig_2}), shows gradual decay of the amplitude away from the FE/FM interface. We clearly see an alternation between maximum and minimum values of the longitudinal magnetization along the both \textit{x}- and \textit{y}-directions ($n=1$ mode). The observed magnetic texture can be explained qualitatively in terms of the noncollinear magnetic order formed in the vicinity of the FE/FM interface. A noncollinear magnetic order occurs due to the dipole-dipole interaction at the FE/FM boundary and enhances at elevated saturation magnetization. The reduction of the saturation magnetization leads to a collinear magnetic order. Our calculations show (not presented here) that the dipole-dipole reservoir dominates over the exchange energy when the saturation magnetization is large, a small saturation magnetization corresponds to the opposite case.  Obviously steering of the interface magnetization $\Delta M_{\mathrm{TiO}_2}$ modifies magnetic order at the FE/FM interface. On the other hand the noncollinear order (because of the transversal components in the coupling term $\Delta M_{\mathrm{TiO}_2}(t)M_{\mathrm{r}}$ where $ M_{\mathrm{r}}$ is the magnetization of the FM part) is very important for the activation of the spin waves. Hence, the features of the emitted spin waves may yield information on the interfacial coupling mechanism, e.g. whether this coupling induces interfacial spin noncollinearity \cite{JiWe14} which is essential for the spin wave emission.

The excitation mechanism of $n=1$ spin wave is as follows: In the upper part of the FE/FM interface magnetization is slightly tilted down respect to the reference direction \textit{x}, while in the bottom part of the FE/FM interface, magnetization is slightly tilted up as is shown in Fig. \ref{fig_1}. With decrease of $ M_{s} $ noncollinear order transforms into collinear. Changes in the upper and lower parts of the FE/FM interface are opposite to each other. Time dependent $\Delta M_{\mathrm{TiO}_2}(t)$ induces the out-of-plane magnetization oscillations in the both upper and lower parts of the quenched area, while the spin wave amplitude is zero in the center ($n=1$ mode spin wave). Besides, the spectrum of the spin waves is monochromatic and contains only main frequency $50$ GHz for both fixed and free magnetization at the left edge (not shown here). No spin waves with other frequency are excited .

\begin{figure}[h]
	\centering
	\includegraphics[width=0.49\textwidth]{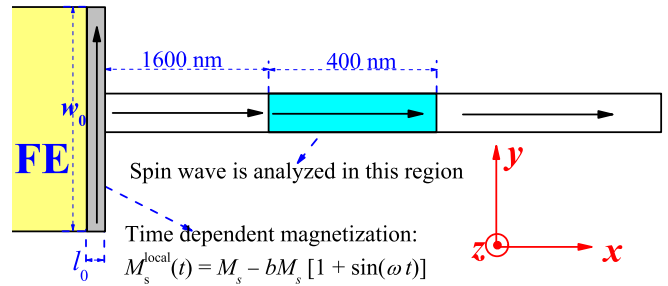}
	\caption{Schematics of the considered two-dimensional $3000\times 30 \times 10$nm$^{3}$-structure with the modified area of the quenched magnetization, where $w_0=210$ nm and $l_0=2$ nm (grey color). The area where the induced spin waves are analyzed is in the middle (400nm-area).}
	\label{fig_3}
\end{figure}

\begin{figure}[h]
	\vspace{2ex}
	\centering
	\includegraphics[width=0.49\textwidth]{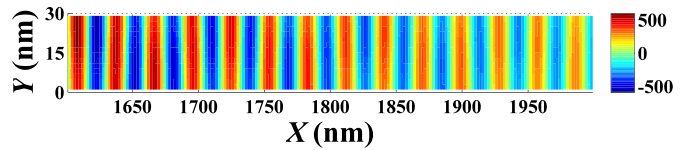}
	\caption{Spatial distribution of the excited spin wave ($n=0$ mode) of the $m_{\mathrm{z}}$-magnetization component for the free quenched magnetization direction ($\omega/(2\pi)=50$ GHz, temperature $T=0$ K, external field $B=0$ T). The modulus of the quenched magnetization alters according $M_{\mathrm{S}} ^{\mathrm{local}}(t)=0.94M_{\mathrm{S}}-0.06M_{\mathrm{S}}\sin(\omega t)$. Unit of the spin-wave profile is A/m.}
	\label{fig_4}
\end{figure}

\begin{figure}[h]
	\vspace{2ex}
	\centering
	\includegraphics[width=0.49\textwidth]{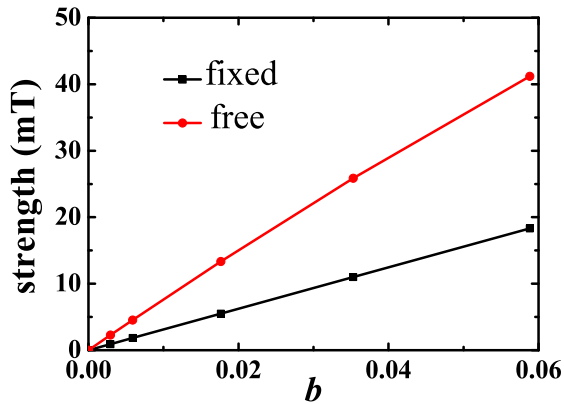}
	\caption{Dependence of the strength of an equivalent magnetic field for inducing spin waves in the FM strip on the amplitude $ b $. The geometry is shown in Fig. \ref{fig_3}}
	\label{fig_5}
\end{figure}

Evidently the spin wave excitation mechanism is related to the noncollinearity. To manipulate noncollinearity and hence manipulate excited spin wave, we suggest a different geometry (Fig. \ref{fig_3}), in which the M-quenched area is broader than the width of the FM stripe. In contrast to the asymmetric noncollinearity, now the noncollinearity along \textit{y} axis is almost uniform and  the excited spin wave is uniform as well ($n=0$ spin wave) see Fig. \ref{fig_4}. The spin wave excitation effect is also stronger. Our calculation has shown that similar effects (exciting  similar spin waves)  are achievable by equivalent local microwave magnetic fields: with the amplitude $41.2$ mT, frequency 50 GHz  and being applied locally in the area of $(0 \leq x \leq 2$ nm). We clearly see that strength of the equivalent local microwave magnetic field linearly increases with the quench amplitude (Fig. \ref{fig_5}).

\begin{figure}[h]
	\vspace{0ex}
	\centering
	\includegraphics[width=0.49\textwidth]{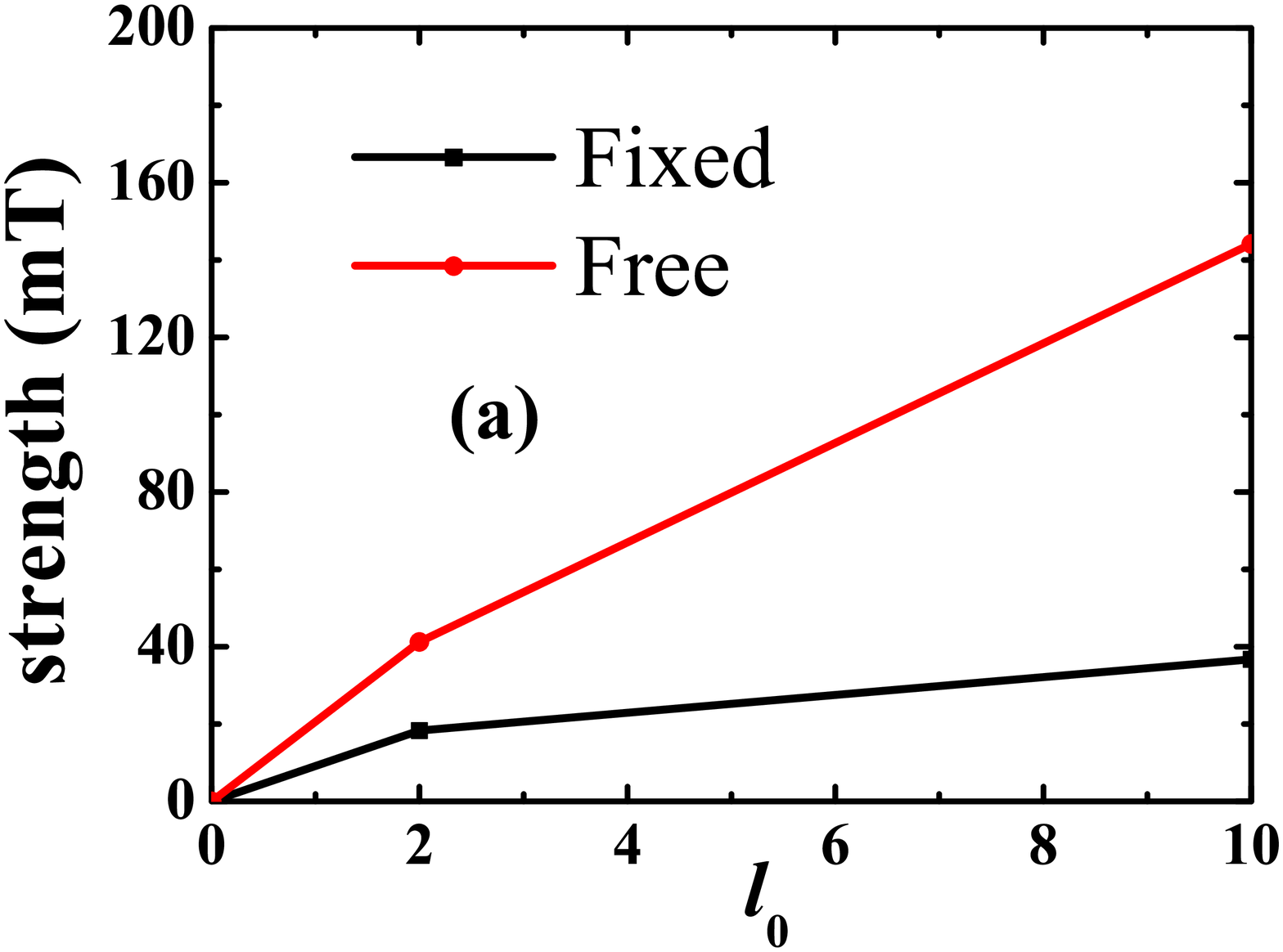}
	\vspace{0ex}
	\includegraphics[width=0.49\textwidth]{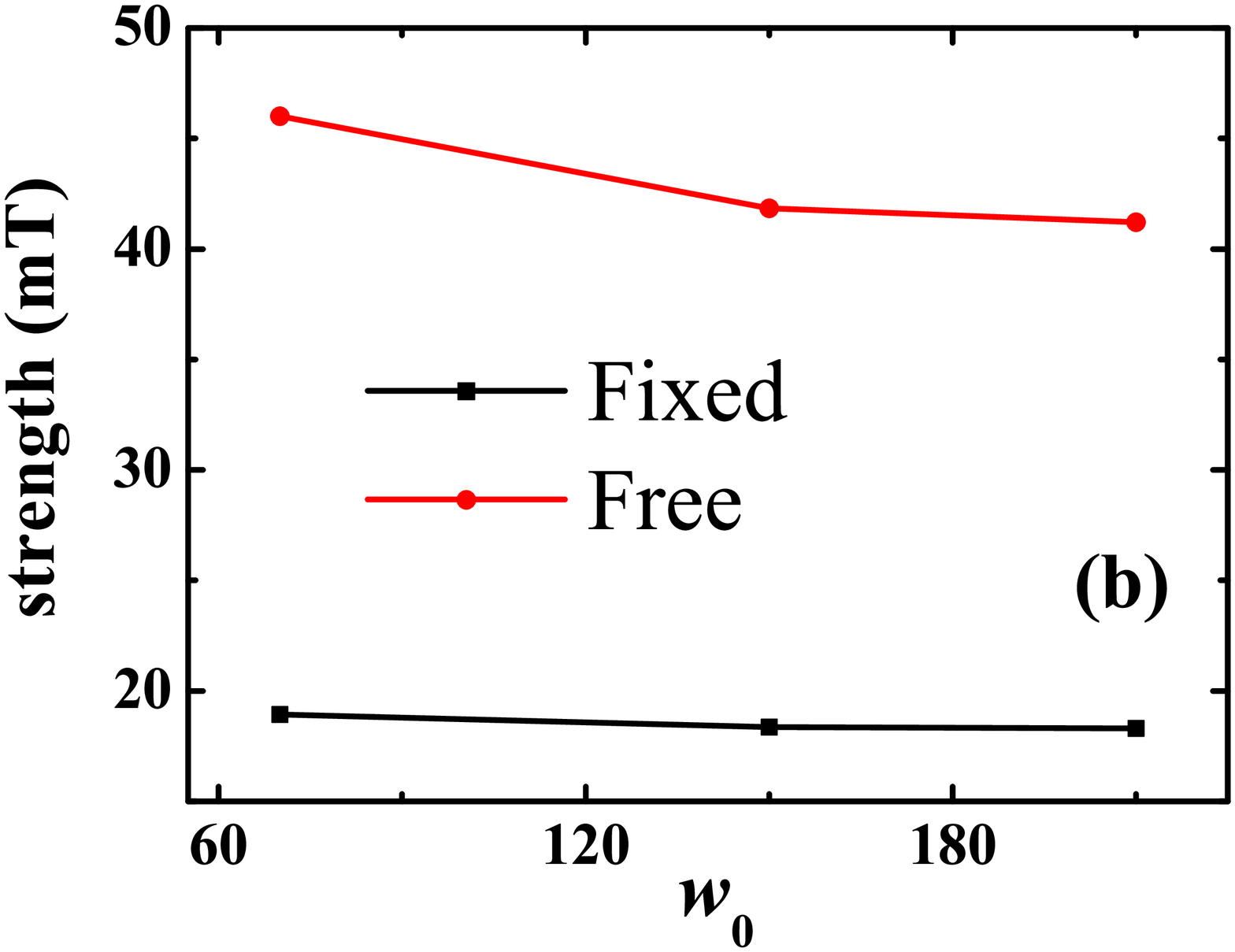}
	\caption{Dependence of the strength of an equivalent magnetic field for generating spin waves in the FM stripe on its length $l_0$ (a) and the width $w_0$ (b), for the geometry shown in Fig. \ref{fig_3}.}
	\label{fig_6}
\end{figure}

For further optimization of the nanostripe's geometry and hence noncollinearity, we explore the dependence of the amplitude of the spin waves (or the strength of the equivalent magnetic field) on the geometry of the quenched area (Fig. \ref{fig_6}). Here, demagnetizing fields play the major role, therefore, if the magnetization direction is free, for significantly high lengths and widths one observes an enhancement of the spin waves' amplitude. Clearly, not the width of the quenched area, but the length of it is more significant (cf. the points for free magnetization, red points in Fig. \ref{fig_6}).

A strong quench (larger $b$)  increases the amplitude of the excited spin waves because of the larger amount of the energy  pumped into the system. The same effect is achieved by  the thickness of the quenched area. An increase of the length $l_{0}$ also increases the number of excited magnetic moments and the amount of the pumped energy. An increase of the width $w_{0}$ presents hwoever no advantage. Because of the geometry of the system, part of the energy is not further transmitted and is wasted.

It is noteworthy that the only $n=0$ spin waves are excited for these geometries here. In addition to the quenched magnetization we also inspected the effect of the decaying magnetization at the FE/FM-interface described in Ref. \cite{JiWe14}. To perform this, the area with the quenched magnetization was modeled according to $M(x,t)= [M_{\mathrm{S}} - b M_{\mathrm{S}} [1 + \sin(\omega t)]] \mathrm{exp}(-x/\gamma)$, where $\gamma=8$~nm is of the order of the spin-diffusion length in Fe. The calculations revealed no sizable changes both for the spin wave spectrum containing the original frequency $50$~GHz only.

\begin{figure}[H]
	\centering
	\includegraphics[width=0.49\textwidth]{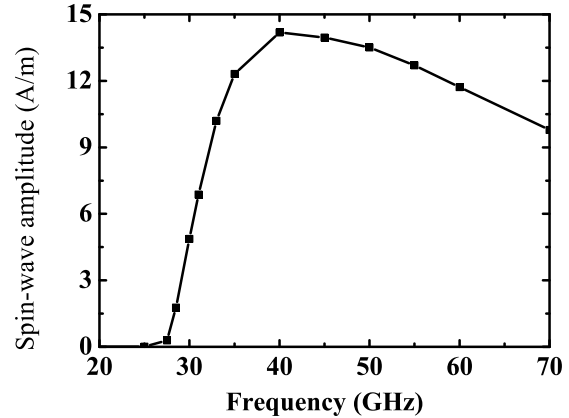}
	\caption{Dependence of the amplitude for triggering spin waves in the FM stripe on the frequency for the geometry shown in Fig. \ref{fig_3}, where $ b = 0.003 $, $w_0=210$ nm and $l_0=2$ nm.}
	\label{fig_7}
\end{figure}

Our simulations  apply to other frequencies, not only 50 GHz. For $w_0=210$ nm and $l_0=2$ nm, the excited spin-wave amplitude as a function of the frequency is shown in Fig. \ref{fig_7}, where $ b = 0.003 $ and the quenched part is fixed. Spin waves having frequencies lower than 28 GHz are prohibited to propagate in the nanostrip, which is in a reasonably good agreement with the estimation of the threshold cutoff frequency (32 GHz). By comparing the Fig. \ref{fig_7} with the local microwave field induced frequency spectrum (not shown here), we find the magnetization excitation effect with $ b = 0.003 $ is equivalent to that of the local microwave field with an amplitude of 1 mT.

\section{Summary}

Summarizing, in a multiferroic composite  consisting of a ferroelectrics coupled to a  ferromagnetic stripe poling the ferroelectric polarization leads to spin wave emission in the ferromagnetic part provided the interfacial magnetoelectric coupling results in an interfacial non-collinear spin order in the ferromagnet. Such coupling mechanisms have indeed been reported theoretically and experimentally \cite{JiWe14,JiWa15}.
  Performing large scale
  micro magnetic simulations for realistic material parameters we demonstrated how the features of the generated spin waves depend on a geometry of the stripe and the coupling at the interface.

%%%%%%%%%%%%%%%%%%%%%%%%%%%%%%%%%%%%%%%%%%%%%%%%%%%%%%%%%%%%%%%%%%%%%%%%%%%%%%%%%%%%%%%%%%

\section{Acknowledgements}
Financial support by the
Deutsche Forschungsgemeinschaft (DFG) through SFB 762,
is gratefully acknowledged.

%%%%%%%%%%%%%%%%%%%%%%%%%%%%%%%%%%%%%%%%%%%%%

%%%%%%%%%%%%%%%%%%%%%%%%%%%%%%%%%%%%%%%%%%

\end{document}